\documentclass[12pt]{JHEP3} 

\usepackage{epsfig,multicol,bbm}

\newcommand\fverb{\setbox\fverbbox=\hbox\bgroup\verb}
\newcommand\fverbdo{\egroup\medskip\noindent%
			\fbox{\unhbox\fverbbox}\ }
\newcommand\fverbit{\egroup\item[\fbox{\unhbox\fverbbox}]}
\newbox\fverbbox
\def\be{\begin{equation}}
\def\ee{\end{equation}}
\def\bs{\begin{subequations}}
\def\es{\end{subequations}}
\def\calA{{\cal A}}
\def\calm{{\cal M}}

\def\calr{{\cal R}}
\def\calrt{\tilde{\cal R}}

\def\ex{\epsilon}
\def\Lx{\Lambda}

\newcommand{\rhot}{\tilde \rho}

\newcommand{\rht}{\tilde{\rho}}

\def\be{\begin{equation}}
\def\ee{\end{equation}}
\def\bs{\begin{subequations}}
\def\es{\end{subequations}}
\newcommand{\een}{\end{subequations}}
\newcommand{\ben}{\begin{subequations}}
\newcommand{\beq}{\begin{eqalignno}}
\newcommand{\eeq}{\end{eqalignno}}

\def\mpl{M_{\rm Pl}}

\title{Effective cosmological equations of induced $f(\calr)$ gravity}

\author{Pantelis S. Apostolopoulos\thanks{Temporary Address: Technological Educational Institute of Messolonghi, Department of Telecommunication Systems 
and Networks, Nafpaktos 30300, Greece}\\ Department of Physics, University of Athens, University Campus, Zographou 157 84, Greece\\ E-mail: \email{papost@phys.uoa.gr}}
\author{Nikolaos Brouzakis\\ Departament de F\'isica, Univeristat Aut\`onoma de Barcelona, 08193 Bellaterra, Barcelona, Spain\\E-mail: \email{nbruzak@ifae.es}}
\author{Nikolaos Tetradis\\ Department of Physics, University of Athens, University Campus, Zographou 157 84, Greece\\ E-mail: \email{ntetrad@phys.uoa.gr}}

\received{\today} 		
\accepted{}		

\abstract{We expand the study of generalized brane cosmologies by allowing for a $f(\calrt)$ gravity term on the brane, with 
$\calrt$ the curvature scalar derived from the induced metric. We also include arbitrary matter components on the brane and 
in the five-dimensional bulk. At low energies, the effect of the bulk on the brane evolution can be described through a 
mirage component, termed generalized dark radiation, in the effective four-dimensional field equations.
Using the covariant formalism, we derive the exact form of these equations.
We also derive an effective conservation equation involving the brane matter and the generalized dark radiation.
At low energies the coupled brane-bulk system has a purely four-dimensional description. 
The applications of the formalism include generalizations of the Starobinsky model and the Dvali-Gabadadze-Porrati cosmology.}

\begin{document}
\section{Introduction}
\setcounter{equation}{0}
The novel characteristics of braneworld cosmological models in the context of the Randall-Sundrum (RS) scenario \cite{rs}
can be understood in terms 
of two basic ingredients: the effective compactification of gravity near a brane
within an Anti-deSitter (AdS) background, and the appearance of a dark (or Weyl or mirage) radiation term 
in the effective equation describing the cosmological evolution \cite{Binetruy:1999ut}.
This term results from the presence of a black hole in the AdS bulk spacetime, whose gravitational field affects the 
brane motion \cite{kraus}. 
 
Starting from this setup, several generalizations have been  
considered, which include modifications of either the 5D bulk spacetime or the brane gravity sector. 
For example, the leading higher-curvature bulk correction in the context of string theory takes the form of a Gauss-Bonnet (GB) term.
The inclusion of such a term results in 
an extension of the conventional braneworld scenario with unusual features \cite{gb1}, such as 
the existence of a self-accelerating branch of the cosmological expansion. 
However, this branch suffers from ghost-like instabilities, which make its physical interpretation problematic
\cite{deser}. 
Radiative corrections can also modify the 4D brane action, 
leading to the Dvali-Gabadadze-Porrati (DGP) model \cite{DGP} and its 
generalization for non-zero brane tension and bulk cosmological constant \cite{inducedrs}. 
The combination of the DGB model with a bulk GB term has also been considered \cite{brown}.
The cosmology of the DGP model has similar properties \cite{acceldgp} and 
pathologies \cite {instability} as the braneworld cosmology with a bulk GB term.
      
Another direction in the generalization of the braneworld scenario is to include matter in the bulk action 
and analyze the effect of its gravitational field on the brane dynamics, as well as its direct interaction with the brane matter \cite{exchange}.
Finding exact and general enough solutions for the bulk geometry is a difficult task.
In the original studies the influence of the bulk matter on the brane cosmological evolution was tractable only in some 
simple cases \cite{ApostTetra1a, exchange2}. This difficulty was overcome by 
employing covariant techniques in order to describe the solution of the braneworld 
equations. 
The covariant approach was used in \cite{Apostolopoulos:2004ic, Apostolopoulos:2006si, Apostolopoulos:2007cr}, where it was found that the 
solution depends on the so called ``comoving mass'' $\mathcal{M}$ of the bulk fluid. This is a function of the brane 
scale factor $R$ and the
proper time $\tau $ on the brane. In the case of an AdS-Schwarzschild bulk it is constant, equal to the black-hole mass.
The form of $\mathcal{M}(R,\tau )$ can be determined only within a specific model of the bulk dynamics. 
Such a model
may involve several bulk fields that possibly interact with the brane, or may employ a description in terms of a bulk cosmological fluid with a certain equation of state. The exact form of $\mathcal{M}(R,\tau )$ is necessary 
for a detailed discussion of the cosmological evolution of the brane. Nevertheless, the general properties of the evolution, such as the occurrence of accelerated expansion, or the rate of energy exchange 
between the bulk and the brane, can be determined by the structure of the resulting covariant equations 
i.e. the Friedmann and Raychaudhuri equations \cite{review}.   The ``comoving mass'' appears in the Friedmann equation 
within a term $\sim \mathcal{M}(R,\tau )/R^4$, characterized as generalized dark radiation. 

Recently, there have been proposals to consider $f(\calr)$ generalizations of the Einstein term in the bulk or brane action.
The inclusion of such a term in the bulk sector \cite {Balcerzak:2010kr, Borzou:2009gn} is possible. 
However, the covariant definition of the junction conditions \cite{israel} on the brane becomes ambiguous 
because of quadratic $\delta$-function contributions to the field equations \cite {Balcerzak:2010kr}. 
Only the Gauss-Bonnet combination of higher derivative terms results in well defined junction conditions \cite{gb1}.
The inclusion of 
higher derivative terms on the brane, represented by a smooth function $f(\calrt)$ of the brane scalar curvature, does not
lead to such difficulties \cite{PapersInducedGravity, Saavedra:2009zz}. 
Moreover, it provides a direct generalisation of the DGP model. 
The purpose of the present paper is to give the set of equations describing the most general formulation of this problem:
brane evolution with higher derivative contributions  resulting from a brane $f(\calrt)$ term, in the presence of arbitrary
matter components on the brane and in the bulk. 

In section 2 we present some general and covariant results regarding the geometrical and dynamical structure 
of the bulk-brane setup. We also give the exact form of the induced Friedmann and Raychaudhuri equations. 
As expected, apart from the higher derivative corrections, these equations contain the generalized dark radiation term.
We also show that there is conservation of energy between the brane matter and the generalized dark radiation, 
generalizing in this way the result of \cite{Apostolopoulos:2004ic, Apostolopoulos:2006si}. 
In section 3 we solve the field equations in a specific coordinate system 
by assuming a static and anisotropic fluid in the bulk. 
We reproduce the Friedmann and conservation equations. In section 4 we show that, at low energies, 
the brane cosmological evolution is that of a 
four-dimensional Universe. The evolution possesses a self-accelerating branch exhibiting three new properties: 
a) an effective cosmological constant, b) an effective negative energy density associated with the generalized dark radiation, and 
c) effective $f(\calrt)$ corrections. Finally in section 5 we draw our conclusions.   

Throughout this paper the following conventions are used: The 5D bulk spacetime manifold is 
endowed with a Lorentzian metric of signature ($-,+,+,+,+$), bulk 5D indices are denoted 
by capital latin letters $A,B,...=0,1,2,...,4$ and greek letters denote brane indices $\alpha ,\beta ,...=0,1,2,3$. 

\section{General covariant results}
\setcounter{equation}{0} 
The effective action we consider has the form 
\begin{equation}
S=\int d^{5}x\sqrt{-g}\left( \Lambda +M^{3}\mathcal{R}+\mathcal{L}%
_{bulk}^{mat}\right) +\int d^{4}x\sqrt{-\tilde{g}}\,\left[ -V+\mathcal{L}%
_{brane}^{mat}+r_{c}M^{3}\,f(\tilde{\mathcal{R}})\right] .  \label{action1}
\end{equation}
In the first integral, $-\Lambda $ is the bulk cosmological constant (we
assume $\Lambda \geq 0$). In the second integral, $V$ is the brane tension, $%
\tilde{g}_{\alpha \beta }$ the induced 4D metric on the brane, $%
\tilde{g}$ its determinant, $\tilde{\mathcal{R}}$ the corresponding
curvature scalar, and $r_{c}$ the characteristic length scale of induced
gravity. The matter contributions are arbitrary, and the effective action
incorporates possible quantum corrections in this sector. We assume,
however, that the corresponding energy-momentum tensor is consistent with
the underlying geometry.

The Einstein field equations (EFE) take the form 
\begin{equation}
G_{AB}=\frac{1}{2M^{3}}\left( T_{AB}^{tot}+\Lambda g_{AB}\right) ,
\label{einstein}
\end{equation}
with the total effective energy-momentum (EM) tensor $T_{AB}^{tot}$ given by 
\begin{equation}
T_{AB}^{tot}=T_{AB}+\delta \left( \eta \right) \tau
_{AB}.  \label{energy-momentum1}
\end{equation}
The term $T_{AB}$ is the bulk matter contribution,
while $\tau _{AB}$ is the contribution from the brane located at $\eta
(x^{A})=0$. The presence of the higher derivative gravity term
 in the brane action, represented by the function $f(\calrt)$, implies that the
tensor $\tau _{\alpha \beta }$ takes the following form \cite
{Saavedra:2009zz} 
\begin{equation}
\tau _{\alpha \beta }=\tilde{T}_{\alpha \beta }-V\tilde{g}_{\alpha \beta
}-2r_{c}M^{3}\hspace{0.15cm}\Sigma _{\alpha \beta }.
\label{modifiedenergymomentum1}
\end{equation}
Here $\tilde{T}_{\alpha \beta }$ is the brane EM tensor
and the function $f(\calrt)$ results in a contribution to the tensor $\tau
_{\alpha \beta }$ proportional to $\Sigma _{\alpha \beta }$, defined as 
\cite{Felice} 
\begin{equation}
\Sigma _{\alpha \beta }=f^{\prime }(\tilde{\mathcal{R}})\tilde{\mathcal{R}}%
_{\alpha \beta }-\frac{1}{2}f(\tilde{\mathcal{R}})\,\tilde{g}_{\alpha \beta
}-\tilde{\nabla}_{\alpha }\tilde{\nabla}_{\beta }f^{\prime }(\tilde{\mathcal{%
R}})+\tilde{g}_{\alpha \beta }{\tilde{\nabla}^{2}}f^{\prime }(\tilde{%
\mathcal{R}}),  \label{sigab}
\end{equation}
where $f^{\prime }(\tilde{\mathcal{R}})=\partial f/\partial \tilde{\mathcal{R%
}}$ and the covariant derivatives are constructed with $\tilde{g}_{\alpha
\beta }$.

In the analysis of cosmological setups, it is natural to restrict our formalism to
geometrical backgrounds with a high degree of symmetry. We make the simplest
symmetry assumption and derive the general form of the solution of the
above equations in the case of a Friedmann-Robertson-Walker (FRW) brane. As a result,
there exist 3D hypersurfaces $\mathcal{D}$, invariant
under a six-dimensional group of isometries. This implies that the surfaces $\mathcal{D}$ 
have constant curvature, parametrized by the constant $k_{c}=0,\pm 1$. 
It should be noted that the assumption of maximal symmetry
of $\mathcal{D}$ implies the existence of a preferred spacelike direction
that represents the local axis of symmetry, with respect to which all the
geometrical, kinematical and dynamical quantities are invariant. This
direction can be chosen in two different ways \cite{Apostolopoulos:2004ic},
which are observer-depended: either w.r.t. the (prolongated) brane observers 
$\tilde{u}^{A}$, for which the spatial direction is represented by the vector
field $n^{A}$ ($n^{A}n_{A}=1$, $\tilde{u}^{A}n_{A}=0$); or w.r.t. the
(comoving) bulk observers $u^{A}$ for which the spacelike vector field is 
$e^{A}$ ($e^{A}e_{A}=1$, $u^{A}e_{A}=0$). The two approaches yield identical results for the cosmological
evolution. Because of the presence of the $\Sigma_{\alpha \beta}$ term in the 
total brane EM tensor, we find it convenient in the present paper to apply the methodology of 
\cite{Apostolopoulos:2007cr}, which is based on the comoving brane observers. In order to simplify the 
analysis, we make the usual assumption of a $Z_2$ symmetry around the brane.

The structure of the bulk-brane setup can be studied using elements from the
theory of spacelike congruences. We introduce the projection tensor \cite
{Apostolopoulos:2007cr} 
\begin{equation}
\Pi _{AB}\equiv g_{AB}+\tilde{u}_{A}\tilde{u}%
_{B}-n_{A}n_{B}=h_{AB}-n_{A}n_{B}  \label{projectiontensor}
\end{equation}
\begin{equation}
\Pi _{A}^{\hspace{0.15cm}A}=3,\hspace{0.1cm}\Pi _{C}^{\hspace{0.15cm}A}\Pi
_{B}^{\hspace{0.15cm}C}=\Pi _{B}^{\hspace{0.15cm}A},\hspace{0.1cm}\Pi _{B}^{%
\hspace{0.15cm}A}n^{B}=\Pi _{B}^{\hspace{0.15cm}A}\tilde{u}^{B}=0,
\label{projectionproperties}
\end{equation}
which is identified with the associated metric of the 3D manifold $\mathcal{D%
}$ (the screen space), normal to the pair $\left\{ \tilde{u}%
^{A},n^{A}\right\} $ at any spacetime event. The first covariant derivatives
of the spacelike vector field $n^{A}$ are decomposed according to 
\begin{equation}
n_{A;B}=\frac{\mathcal{\vartheta }}{3}\Pi _{AB}-\dot{n}_{A}\tilde{u}%
_{B}+n_{A}^{\prime }n_{B}=K_{AB}+n_{A}^{\prime }n_{B}.
\label{derivativedecomposition1}
\end{equation}
Here 
\begin{equation}
\mathcal{\vartheta }=n_{A;B}\Pi ^{AB}=n_{\hspace{0.2cm};A}^{A}+\dot{n}_{A}\tilde{u}^{A}  \label{expansion}
\end{equation}
is the rate of the surface expansion $\mathcal{D}$, and we have used the
notation 
\begin{equation}
P_{A...}^{\prime }\equiv P_{A...;L}n^{L},\hspace{0.7cm}\dot{P}_{A...}\equiv
P_{A...;L}\tilde{u}^{L}  \label{ederivativedefinition}
\end{equation}
for the directional derivative along the vector fields $n^{A},\tilde{u}^{A}$
of any scalar or tensorial quantity. In addition 
$K_{AB}=({\mathcal{\vartheta }}/{3})\Pi _{AB}-\dot{n}_{A}\tilde{u}_{B}$ 
corresponds to the 
extrinsic curvature of the timelike hypersurfaces normal to $n^{A}$. 

The
definition of the overall expansion $\mathcal{\vartheta }$ of the spacelike
congruence implies 
\begin{equation}
\mathcal{\vartheta }=\Pi ^{AB}K_{AB}\equiv n_{\hspace{0.15cm}\parallel
A}^{A}\equiv 3\frac{\ell ^{\prime }}{\ell },  \label{lengthscale1}
\end{equation}
where $\ell $ can be interpeted as the average length scale of $\mathcal{D}$. 
For example, in the spherically symmetric case $k=1$ it
represents the radius of the spheres $\mathcal{D}$. 
The temporal change of $\ell $ is controlled by the expansion rate of the timelike
congruence as measured in the screen space $\mathcal{D}$, namely 
\begin{equation}
\Pi ^{AB}\tilde{u}_{A;B}\equiv \tilde{u}_{\hspace{0.15cm}\parallel A}^{A}=3\frac{%
\dot{\ell}}{\ell }.  \label{lengthscale2}
\end{equation}
Therefore, the length $\ell $ completely determines the volume of $\mathcal{D}$.
For on brane considerations, it is identified with the scale factor
of the brane Universe. Employing the Gauss-Codacci equations for the
spacelike surfaces $\mathcal{D}$, it can be shown that \cite{Apostolopoulos:2007cr} 
\begin{equation}
\frac{k}{\ell ^{2}}=-\frac{1}{3}\left( \mathcal{E}+\frac{1}{2}%
G_{AB}g^{AB}-2G_{\perp }\right) -\left( \frac{\dot{\ell}}{\ell }\right)
^{2}+\left( \frac{\ell ^{\prime }}{\ell }\right) ^{2}
\label{3ScalarCurvature2}
\end{equation}
\begin{equation}
\left( \frac{\dot{\ell}}{\ell }\right) ^{\cdot }+\left( \frac{\dot{\ell}}{%
\ell }\right) ^{2}=\frac{1}{3}\left( \mathcal{E}+\frac{1}{2}%
G_{AB}g^{AB}-2G_{\perp }\right)-\frac{\ell ^{\prime }}{\ell }\dot{n}_{A}\tilde{u}_{B}-%
\frac{1}{3}G_{AB}n^{A}n^{B}.  \label{Raychaudhuri1}
\end{equation}
The generalized dark radiation term $\mathcal{E}=C_{ACBD}\tilde{u}^{A}n^{C}%
\tilde{u}^{B}n^{D}=C_{ACBD}u^{A}e^{C}u^{B}e^{D}$ is the spatial eigenvalue
of the electric part of the 5D Weyl tensor, and $3G_{\perp }\equiv \Pi
^{AB}G_{AB}$. Equation (\ref{3ScalarCurvature2}) shows how the scalar
curvature of the 3D space $\mathcal{D}$ is affected by the kinematics and the
dynamical (when the 5D EFE are employed) content of the spacetime.
At the location of the brane, equations (\ref{3ScalarCurvature2}) and (\ref{Raychaudhuri1}) correspond
to the effective Friedmann and Raychaudhuri equations of the FRW
brane.

The matter content of the bulk is described by the EM tensor $T_{AB}$, 
which can be written in the usual way with respect
to the bulk observers $u^{A}$: 
\begin{equation}
T_{AB}=\rho u_{A}u_{B}+ph_{AB}+2q_{(A}u_{B)}+\pi _{AB},
\label{energy-decomp2}
\end{equation}
where 
\begin{eqnarray}
\rho  &=&T_{AB}u^{A}u^{B},
\,\,\,\,
p=\frac{1}{4}T_{AB}h^{AB},
\,\,\,\,
q_{A}=-h_{A}^{C}T_{CD}
u^{D}, \\
&& \nonumber \\
\pi _{AB} &=&h_{A}^{C}h_{B}^{D}T_{CD}-\frac{1}{4}
h^{CD}T_{CD}\,h_{AB}.  \label{dynam-quantities}
\end{eqnarray}
The first term in the rhs of equations (\ref{3ScalarCurvature2}) and (\ref
{Raychaudhuri1}) can be determined through the full 5D EFE (\ref{einstein})
and the bulk EM tensor (\ref{energy-decomp2}). The result is 
\cite{Apostolopoulos:2004ic, Apostolopoulos:2006si}
\begin{equation}
\frac{k}{\ell ^{2}}=\frac{\mathcal{M}}{6M^{3}\pi ^{2}\ell ^{4}}-\frac{%
\Lambda }{12M^{3}}-\left( \frac{\dot{\ell}}{\ell }\right) ^{2}+\left( \frac{%
\ell ^{\prime }}{\ell }\right) ^{2}  \label{Fried1}
\end{equation}
\begin{equation}
\left( \frac{\dot{\ell}}{\ell }\right) ^{\cdot }+\left( \frac{\dot{\ell}}{%
\ell }\right) ^{2}=-\frac{\ell ^{\prime }}{\ell }\dot{n}_{A}\tilde{u}^{A}-%
\frac{\mathcal{M}}{6M^{3}\pi ^{2}\ell ^{4}}-\frac{\Lambda }{12M^{3}}-\frac{1%
}{6M^{3}}\bar{p}_{\parallel },  \label{Raychaudhuri2}
\end{equation}
where $\bar{p}_{\parallel }=T_{AB}n^{A}n^{B}$ is the
pressure along the direction of the preferred spacelike vector field $n^A$, and $%
\mathcal{M}$ is the ``comoving mass" of the bulk fluid, satisfying 
\begin{equation}
\left( \mathcal{M}-\mathcal{M}_{0}\right) ^{\prime }=2\pi ^{2}\rho \ell
^{3}\ell ^{\prime }.  \label{mass-function}
\end{equation}
Only in the spherically symmetric case ($k=1$) the ``comoving mass" $\mathcal{M}
$ has the usual physical interpretation as the effective gravitational mass
contained within a sphere with radius $\ell $.
However, we shall refer to $\mathcal{M}$ as the ``comoving mass" for all
geometries of the hypersurfaces $\mathcal{D}$. The integration constant
$\mathcal{M}_{0}$ in equation (\ref{mass-function}) can be interpreted as
the mass of a black hole at $\ell _{0}=0$.

The covariant form of the junction conditions for braneworld models has been
derived in \cite{maeda} and involves the projected (perpendicular to $n^A$) part of the extrinsic
curvature 
\begin{equation}
K_{\alpha \beta }=-{\frac{1}{4M^{3}}}\left( \tau _{\alpha \beta }-{\frac{1}{3%
}}\tau g_{\alpha \beta }\right).   \label{Junction1}
\end{equation}
Equation (\ref{Junction1}) can be used in order to determine the
discontinuous quantities ${\ell ^{\prime }}/{\ell }$ and 
$\dot{n}_{A}\tilde{u}^{A}$ by projecting along $\Pi ^{\alpha \beta }$ and 
$\tilde{u}^{\alpha }\tilde{u}^{\beta }-2\Pi ^{\alpha \beta }/3$ respectively.
In addition, assuming a perfect fluid matter configuration on the brane, the
brane EM tensor can be written in terms of the energy density $\tilde{%
\rho}$ and the isotropic pressure $\tilde{p}$ as 
\begin{equation}
\tilde{T}_{\alpha \beta }=\tilde{\rho}\tilde{u}_{\alpha }\tilde{u}_{\beta }+%
\tilde{p}\tilde{h}_{\alpha \beta }.  \label{braneenergymomentum}
\end{equation}
At the location of the brane, we deduce that 
\begin{equation}
\frac{\ell ^{\prime }}{\ell }=-\frac{1}{12M^{3}}\left\{ V+\tilde{\rho}%
-2r_{c}M^{3}\left[ 3f^{\prime }\left( H^{2}+\frac{k_{c}}{R^{2}}\right) -%
\frac{1}{2}\left( \tilde{\mathcal{R}}f^{\prime }-f\right) +3H\dot{\tilde{%
\mathcal{R}}}f^{\prime \prime }\right] \right\}   \label{lPrime1}
\end{equation}
\begin{equation}
\dot{n}_{A}\tilde{u}^{A}=-\frac{1}{12M^{3}}\left\{ 3\tilde{p}+2\tilde{\rho}%
-V+2r_{c}M^{3}\left[ 9f^{\prime }\frac{\ddot{R}}{R}+3\left( f^{\prime
}\right) ^{\cdot \cdot }-\tilde{\mathcal{R}}f^{\prime }+f\right] \right\} .
\label{SpatialVectorDot1}
\end{equation}
We recall that for on-brane
considerations, $\ell =R$, with $R$ the scale factor of the brane. 
The directional derivative denoted by a dot (defined in equation \ref{ederivativedefinition}) becomes 
a derivative with respect to the proper time on the brane. 
As a result, $H={\dot{R}}/{R}$ is the rate of the
cosmological expansion. 
Also, $f$, $f'$, $f''$ are considered functions of $\calrt=6(\dot{H}+2H^2+k_c/R^2)$.
The sign of the directional derivative $\ell'$ determines the way the brane is embedded in the bulk spacetime. 
For $r_{c}=0$, a negative sign for $\ell'$ results in localization of the low energy gravitons near the brane.
It is apparent from equation (\ref{lPrime1}) that $\ell'<0$ corresponds to a 
brane with positive tension, a property that guarantees stability under small
perturbations. The
self-accelerating branch of the DGP model \cite{DGP, acceldgp} at late times has 
$V=0$, $f(\calrt)=\calrt$, $\tilde{\rho},a^{-1}\rightarrow 0$ and $H^{2}\sim 1/r_{c}$. It
is then apparent from (\ref{lPrime1}) that $\ell'>0$ in this case.  
This branch is known to have ghost-like instabilities 
\cite{instability}.

Using equations (\ref{Fried1}) and (\ref{lPrime1}), the effective
Friedmann equation becomes 
\begin{eqnarray}
&&H^{2}+\frac{k_{c}}{R^{2}}+\frac{\Lambda }{12M^{3}}-\frac{\mathcal{M}%
(R,\tau )}{6\pi ^{2}M^{3}R^{4}}=  \nonumber \\
&&\frac{1}{144M^{6}}\left\{ V+\tilde{\rho}-2r_{c}M^{3}\left[ 3f^{\prime
}\left( H^{2}+\frac{k_{c}}{R^{2}}\right) -\frac{1}{2}\left( \tilde{\mathcal{R%
}}f^{\prime }-f\right) +3H\dot{\tilde{\mathcal{R}}}f^{\prime \prime }\right]
\right\} ^{2}. \label{Friedmann}\\
&& \nonumber \end{eqnarray}
This equation represents the most general form of the brane
cosmological evolution in the presence of bulk and brane matter and
higher derivatives terms in the brane action. The bulk affects the brane evolution through the
``comoving mass'' $\calm$, which is a function of the scale factor $R$, but may also have an explicit dependence on
the proper time on the brane.  
The effective Friedmann equation (\ref{Friedmann})
results from squaring $\ell'$. It includes both the
normal and self-accelerating branches of the DGP model.

Throughout this paper we assume the standard tuning to zero of the 
effective cosmological constant in the Randall-Sundrum scenario \cite{rs}.
This is achieved if the bulk cosmological constant $-\Lx$ and the 
brane tension $V$ are related through $\Lx=V^2/(12M^3)$.
We also define the energy scale $k=V/(12M^3)=\left[\Lx/(12M^3)\right]^{1/2}$. 
The term 
\begin{equation}
\rht_d=\frac{\calm(R,\tau)}{k \pi^2 R^4 }
\label{darkradi1} \end{equation}
is the effective energy density of the generalized dark radiation \cite{review} and reflects the 
influence of the bulk matter on the brane evolution. 

The Raychaudhuri equation (\ref{Raychaudhuri2}) can be used in
order to derive the effective conservation equation between the brane and
the bulk. Taking the derivative of the Friedmann equation 
(\ref{Friedmann}) along $\tilde{u}^{\alpha }$, combining the result with
(\ref{Raychaudhuri2}), and using (\ref{Fried1}), (\ref{lPrime1})-(\ref{SpatialVectorDot1}), 
we find the remarkably simple effective
conservation equation
\begin{eqnarray}
&&\dot{ \tilde{\rho}}+3H\left( \tilde{\rho}+\tilde{p}\right)  =
\nonumber \\
&&-\epsilon_2 k
\left[\dot{\rht}_d+H\left( 4\rht_d+\frac{2}{k}\bar{p}_{\parallel }\right) \right]
\left( H^{2}+\frac{k_{c}}{R^{2}}+k^2 
-\frac{k}{6M^3}\rht_d\right) ^{-1/2}.
\label{ConservationEquation1}
\end{eqnarray}
The two possible values $\epsilon_2=1$ or $-1$ correspond to the two embeddings of the brane
in the bulk spacetime.  For $r_c=0$, these require a brane tension $V>0$ or $V<0$, respectively.

\section{A case study: Anisotropic bulk fluid}
In the previous section we derived the most general form of the Friedmann 
and Raychaudhuri equations for an arbitrary bulk matter configuration. However, 
 it is instructive to analyze a class of cases in which the problem is tractable
in specific coordinate systems. We assume that for a certain observer the
bulk content can be described as a static fluid. This assumption allows the
possibility of an arbitrary number of fields and relies only on the
existence of an observer comoving with the bulk matter. Clearly, important
physical situations, such as those that involve the propagation of
electromagnetic or gravitational radiation, are excluded by our assumption.
However, many interesting backgrounds, including generalized black-hole
ones, are allowed.

In order for the embedding of a cosmological 3-brane to be possible, the
spatial part of the metric must include a 3-space of constant curvature. The
resulting metric can be cast in the form 
\begin{equation}
ds^{2}=-n^{2}(r)dt^{2}+r^{2}d\Omega _{k}^{2}+b^{2}(r)dr^{2}.  \label{metric}
\end{equation}%
The lhs of the EFE\ (\ref{einstein}) take the form 
\begin{eqnarray}
{{G}}_{~0}^{\,0} &=&\frac{3}{b^{2}}\frac{1}{r}\left( \frac{1}{r}-\frac{%
b^{\prime }}{b}\right) -\frac{3k_c}{r^{2}} \label{ein00} \\
{{G}}_{~j}^{\,i} &=&\frac{1}{b^{2}}\left[ \frac{1}{r}\left( \frac{1}{r}+2%
\frac{n^{\prime }}{n}\right) -\frac{b^{\prime }}{b}\left( \frac{n^{\prime }}{%
n}+2\frac{1}{r}\right) +\frac{n^{\prime \prime }}{n}\right] -\frac{k_c}{r^{2}}
 \label{einij} \\
{{G}}_{~4}^{\,4} &=&\frac{3}{b^{2}}\frac{1}{r}\left( \frac{1}{r}+\frac{%
n^{\prime }}{n}\right) -\frac{3k_c}{r^{2}},  \label{ein44}
\end{eqnarray}%
where here the prime denotes a derivative with respect to $r$.

The general form of the bulk EM tensor consistent with the above geometric
setup is 
\begin{equation}
T^A_{~B}=\mathrm{diag} \left( -\rho, \mathrm{p}, 
\mathrm{p}, \mathrm{p}, p \right),  \label{PerfectFluidEM1}
\end{equation}%
with the two pressures $\mathrm{p}$, $p$ not equal unless the bulk matter
can be interpreted as a perfect fluid. The 00 component of (\ref{einstein})
gives 
\begin{equation}
\left( \frac{r^{2}}{b^{2}} \right)^{\prime }
=2k_c r+\frac{1}{3M^{3}}r^{3}(\Lambda -\rho ),  \label{ijre}
\end{equation}
whereas the combination of the 00 and 44 components results in 
\begin{equation}
\frac{(bn)^{\prime }}{bn} =\frac{1}{6M^{3}}\,b^{2}r\,(\rho +p).  \label{extra1}
\end{equation}%
The conservation of the bulk EM tensor can be written in the
form 
\begin{equation}
\frac{p^{\prime }}{\rho +p}+ \frac{3(p-\mathrm{p})}{r(\rho+p)} =
-\frac{1}{6M^3}
 (p+\Lambda)r b^{2}+\frac{1-k_c b^{2}}{r}.  \label{Conservation1}
\end{equation}%
Because of the Bianchi identities, the set (\ref{ijre})-(\ref{Conservation1}%
) completely describes the solution.

Integrating (\ref{ijre}) we find 
\begin{equation}
\frac{r^{2}}{b^{2}}=k_cr^{2}+%
\frac{\Lambda r^{4}}{12M^{3}}-\frac{\mathcal{M}(r)}{6\pi ^{2}M^{3}},  \label{sol1}
\end{equation}%
where $\mathcal{M}(r)$ satisfies 
\begin{equation}
\frac{d\mathcal{M}}{dr}=2\pi ^{2}r^{3}\rho  \label{sol1b}
\end{equation}%
and corresponds to the ``comoving mass" of the bulk fluid.

In order to analyze the cosmological evolution of the brane, we employ the
Gaussian normal coordinate system in which the metric takes the form 
\begin{equation}
ds^{2}=-m^{2}(\tau ,\eta )d\tau ^{2}+a^{2}(\tau ,\eta )d\Omega
_{k}^{2}+d\eta ^{2},  \label{sx2.1ex}
\end{equation}%
with $m(\tau ,\eta =0)=1$. Through an appropriate coordinate transformation 
\begin{equation}
t=t({\tau },\eta ),\qquad r=r({\tau },\eta )  \label{sx2.4}
\end{equation}%
the metric (\ref{metric}) can be written in the form of equation (\ref%
{sx2.1ex}). We define $R({\tau })=a({\tau },\eta =0)$. In the system of
coordinates $(t,r)$ of equation (\ref{metric}) the brane is moving, as it is
located at $r=R({\tau })$. Hence \cite{ApostTetra1a} 
\begin{eqnarray}
\frac{\partial t}{\partial \tau } &=&\frac{1}{n(R)}\left[ b^{2}(R)\dot{R}%
^{2}+1\right] ^{1/2}  \label{tr1} \\
\frac{\partial t}{\partial \eta } &=&-\epsilon _{2}\frac{b(R)}{n(R)}\dot{R}
\label{tr2} \\
\frac{\partial a}{\partial \tau } &=&\dot{R}  \label{tr3} \\
\frac{\partial a}{\partial \eta } &=&-\epsilon _{2}\frac{1}{b(R)}\left[
b^{2}(R)\dot{R}^{2}+1\right] ^{1/2},  \label{tr4}
\end{eqnarray}%
where the dot denotes a derivative with respect to proper time and $\epsilon
_{2}=\pm 1$. The $\eta $-derivatives are evaluated for $\eta =0^{+}$. The
value of $\epsilon _{2}$ determines the way the brane is embedded in the
bulk space. As we have mentioned earlier, we impose a $Z_{2}$-symmetry
around the brane. We consider a matter configuration that is consistent with the
solution of the EFE in an
infinite bulk before the brane embedding. When the brane is included, only the solution in 
half of the space and its mirror image are employed. 
The value of $\epsilon _{2}$ determines which half-space
is used. A negative sign in the rhs of equation (\ref{tr4}) 
means that $r$ decreases away from the brane. In the absence of induced
gravity, the brane has positive tension. The configuration is
stable under small perturbations and the massless graviton is localized near
the brane.

The bulk EM tensor at the location of the brane in the
coordinate system $({\tau },\eta )$ is  (no summation over repeated indices) 
\begin{eqnarray}
T_{00} &=&\rho (R)+\left[ \rho (R)+p(R)\right]
b^{2}(R)\dot{R}^{2}  \label{t00} \\
T_{ii} &=&R^{2}\mathrm{p}(R)\hspace{0.3cm}  \label{t11} \\
T_{44} &=&p(R)+\left[ \rho (R)+p(R)\right] b^{2}(R)%
\dot{R}^{2}  \label{t44} \\
T_{04} &=&\epsilon_2 b(R)\dot{R}\left[ b^{2}(R)\dot{R}%
^{2}+1\right] ^{1/2}\left[ \rho (R)+p(R)\right] .  \label{t04}
\end{eqnarray}%
The sign of $T_{04}$ indicates whether a brane
observer detects inflow or outflow of energy. This sign is determined by the
value of $\epsilon_2$ and the rate of expansion (or contraction) $\dot{R}/R$ of
the scale factor on the brane. 

The lhs of (\ref{einstein}) near the brane $(\eta \rightarrow 0^{\pm })$
take the form (no summation over repeated indices) 
\begin{eqnarray}
{{G}}_{~0}^{\,0} &=&\frac{3a^{\prime }{}^{2}}{a^{2}}
-\frac{3\dot{a}^{2}}{a^{2}}
+\frac{3a^{\prime \prime }}{a}
-\frac{3k_c}{a^{2}}  \label{Gaussein00} \\
{{G}}_{~i}^{\,i} &=&\frac{\left( a^{\prime }\right) ^{2}}{a^{2}}
+\frac{%
2m^{\prime }a^{\prime }}{a}
-\frac{\dot{a}^{2}}{a^{2}}
+\frac{2a^{\prime \prime }}{a}+m^{\prime \prime }-\frac{2\ddot{a}}{a}-\frac{k_c}{a^{2}}
\label{Gausseinij} \\
{{G}}_{~4}^{\,4} &=&-\frac{12am^{\prime }\left( a^{\prime }\right) ^{3}}{%
a^{3}}+\frac{%
3\left( a^{\prime }\right) ^{2}}{a^{2}}
+\frac{3m^{\prime }a^{\prime }}{a}-\frac{3%
\dot{a}^{2}}{a^{2}}-\frac{3\ddot{a}}{a}-\frac{3k_c}{a^{2}}  \label{Gaussein44} \\
{{G}}_{~4}^{\,0} &=&-\frac{3m^{\prime }\dot{a}}{a}+\frac{3\dot{a}^{\prime }}{a},
\label{Gaussein40}
\end{eqnarray}%
with a prime now denoting a derivative with respect to $\eta $.

We consider a brane Universe containing a perfect fluid with an
EM tensor 
\begin{equation}
T_{AB}=\delta (\eta )a^{2}(\tau ,\eta )\mathrm{diag}%
\left[ \frac{m^{2}(\tau ,\eta )}{a^{2}(\tau ,\eta )}\tilde{\rho},\tilde{p},%
\tilde{p},\tilde{p},0\right] .  \label{branet}
\end{equation}%
Integrating the 00 component of (\ref{einstein}) on a small $\eta $
interval around the brane and using (\ref{modifiedenergymomentum1}) we
obtain 
\begin{eqnarray}
&&\frac{a_{+}^{\prime }}{a}
=-\frac{1}{12M^{3}}\left \{ V+\tilde{\rho}-2r_{c}M^{3}
\left[ 3f' \left( H^{2}+\frac{k_c}{a^{2}}\right) 
 -\frac{1}{2}\left(\calrt f'-f\right ] 
+3 H \dot{\calrt}f''\right) \right\},  
\nonumber \\
&&
 \label{JunctionGauss1} 
\end{eqnarray}%
where $H=\dot{R}/R$ and $f$, $f'$, $f''$ are considered functions of $\calrt=6(\dot{H}+2H^2+k_c/R^2)$.  
From (\ref{JunctionGauss1}) and (\ref{tr4}) it is straightforward to derive
the effective Friedmann equation (\ref{Friedmann}).

In the low-energy limit ($\tilde{\rho},H,a^{-1}\rightarrow 0$) our choice of
sign for $a_{+}^{\prime }$ in (\ref{tr4}) must be consistent with the rhs of
(\ref{JunctionGauss1}). For example, for $r_{c}=0$ a negative sign ($%
\epsilon _{2}=1$) for $a_{+}^{\prime }$ is consistent with the negative sign
in the rhs of (\ref{JunctionGauss1}) only if $V>0$. The
self-accelerating branch of the DGP model \cite{DGP, acceldgp} at late times has 
$V=0$, $f(\calrt)=\calrt$, $\tilde{\rho},a^{-1}\rightarrow 0$ and $H^{2}\sim 1/r_{c}$. It
is then apparent from (\ref{JunctionGauss1}) that $a_{+}^{\prime }>0$ ($%
\epsilon _{2}=-1$). The effective Friedmann equation (\ref{Friedmann})
results from squaring $a_{+}^{\prime }$. As a result, it includes both the
normal and self-accelerating branches of the DGP model.

The effective Friedmann equation can be complemented by a 
conservation equation for the energy densities that drive the expansion. 
In order to obtain this equation we 
integrate the 00 and ii components of (\ref{einstein}) on a small $\eta $
interval around the brane. This gives (no summation is implied)
\begin{eqnarray}
&&\frac{a'_+}{a}=\frac{1}{12M^3}\tau^0_{~0}
\nonumber \\
&&m'_+=\frac{1}{4M^3}\tau^i_{~i}-\frac{1}{6M^3}\tau^0_{~0}
\label{amp} \end{eqnarray}
at the location of the brane. From equation (\ref{Gaussein40}) we now find 
\begin{equation}
G^0_{~4}=\frac{1}{4M^3}\left[ \dot{\tau}^0_{~0} +3 H (\tau^0_{~0}-\tau^i_{~i}) \right]
\label{g04br} \end{equation}
at the brane. 
The modified energy momentum tensor of the brane, given by equation (\ref{modifiedenergymomentum1}),
contains a piece $\Sigma_{\alpha\beta}$ that arises from the induced gravity term in the action (\ref{action1}).
We have verified that $\Sigma_{\alpha\beta}$ is conserved.
Therefore, the rhs of equation (\ref{g04br}) receives contributions only from the density and pressure of the brane matter, so 
that 
\begin{equation}
G^0_{~4}=-\frac{1}{4M^3}\left[ \dot{\rhot} +3 H (\rhot+\tilde{p}) \right].
\label{g04brnew} \end{equation}

The bulk energy momentum at the location of the brane, as measured by an observer comoving with the brane, 
is given by equations (\ref{t00})-(\ref{t04}).
Solving equation (\ref{t44}) for $p(R)$ and substituting in equation (\ref{t04}) gives
\begin{equation}
T^0_{~4}=-\epsilon_2\frac{\dot{R}}{R}
\left[T_{44}+\rho(R)\right]
\left[ \frac{1}{R^2\,b^2(R)}+\frac{\dot{R}^2}{R^2}\right]^{-1/2}.
\label{t04new} \end{equation}
The energy density $\rho$ in the above expression corresponds to the bulk energy density as measured by 
a bulk observer. It is more convenient to express it terms of the energy density of dark radiation 
(\ref{darkradi1}). Making use of equation (\ref{sol1b}) we find
\be
\rho(R)=\frac{k}{2H}\left(\dot{\rhot}_d+4H\rhot_d \right).
\label{darkra} \ee
The component $T_{44}$ determines the bulk pressure as measured by a comoving brane observer.
It is equal to the quantity $\bar{p}_{\parallel}$, covariantly defined below equation (\ref{Raychaudhuri2}). 
We define an effective pressure term for the dark radiation through the relation
\begin{equation}
\tilde{p}_d=\frac{\rhot_d}{3}+\frac{2}{3k}T_{44}. 
\label{darkpr} \end{equation}
Making use of these definitions, the 04 of the EFE at the location of the brane, and the solution of equation (\ref{sol1}) for $b(R)$,
we find the conservation equation
\begin{eqnarray}
&&-\left[ \dot{\rhot} +3 H (\rhot+\tilde{p}) \right]=
\nonumber \\
&&\epsilon_2
k\left(H^2+\frac{k_c}{R^2}+k^2-\frac{\calm}{6\pi^2M^3R^4} \right)^{-1/2}
\left[ \dot{\rhot}_d +3 H (\rhot_d+\tilde{p}_d) \right].
\label{conss} \end{eqnarray}

\section{Effective equations at low energies}

The main equations that have resulted from our study are the 
effective Friedmann equation (\ref{Friedmann}) and the conservation equation (\ref{ConservationEquation1}).
The Friedmann equation can also be written as 
\begin{eqnarray}
\frac{r_c^2}{2}\left( H^2 + \frac{k_c}{R^2}  \right)
&&=1+ kr_c + kr_c \frac{\rht}{V} + kr_c \frac{\calA}{V} 
\nonumber \\
&&- \ex_2 \left[ 
(1+kr_c)^2+ 2kr_c \frac{\rht}{V} + 2 kr_c \frac{\calA}{V} 
-2 (kr_c)^2\frac{\rht_d}{V}
\right]^{1/2}.
\label{friedm2} \end{eqnarray}
 The term
\begin{equation}
\calA=-6 r_c M^3 \left[ 
(f'-1)\left( H^2 + \frac{k_c}{R^2}  \right)-\frac{1}{6}\left(\calrt f'-f\right)+H\dot{\calrt}f''
\right]
\label{calaa} \end{equation}
incorporates the effects of the deviation of $f(\calrt)$ from the standard Einstein form $f=\calrt$.

For $\rhot, \rhot_d, \calA \ll V$, keeping only the linear terms, we obtain 
\begin{eqnarray}
&&H^2+\frac{k_c}{R^2}+ \frac{1-\ex_2+kr_c}{1+kr_c}\left[ 
(f'-1)\left( H^2 + \frac{k_c}{R^2}  \right)-\frac{1}{6}\left(\calrt f'-f\right)+H\dot{\calrt}f''
\right]=
\nonumber \\
&&\frac{2(1-\ex_2)(1+kr_c)}{r_c^2}
+\frac{1-\ex_2+kr_c}{kr_c(1+kr_c)}\frac{\rht}{6\left(M^3/k\right)}
+\frac{\ex_2}{1+kr_c}\frac{\rht_d}{6\left(M^3/k\right)},
\label{friedm3} \end{eqnarray}
with $\calrt=6(\dot{H}+2H^2+k_c/R^2)$.

For $\ex_2=1$ we have 
\begin{eqnarray}
&&H^2+\frac{k_c}{R^2}+ \frac{kr_c}{1+kr_c}\left[ 
(f'-1)\left( H^2 + \frac{k_c}{R^2}  \right)-\frac{1}{6}\left(\calrt f'-f\right)+H\dot{\calrt}f''
\right]=
\nonumber \\
&&
\frac{1}{6\mpl^2}\left(\rht+\rht_d \right),
\label{friedm4} \end{eqnarray}
with $\mpl^2=M^3(r_c+1/k)$.
For $kr_c \gg 1$, the lhs takes the exact form it would have for four-dimensional $f(\calrt)$ theories.
The sources, including the energy
density of the generalized dark radiation contribute linearly in the rhs of the same equation.
For $kr_c \ll 1$, the lhs of the Friedmann equation takes the conventional form, as expected for the 
low-energy limit of RS cosmology. 

For $\ex_2=-1$ and $kr_c \gg 1$ we obtain
\begin{equation}
f'\left( H^2 + \frac{k_c}{R^2}  \right)-\frac{1}{6}\left(\calrt f'-f\right)+H\dot{\calrt}f''
=
\frac{4k}{r_c}+
\frac{1}{6\mpl^2}\left(\rht-\rht_d \right),
\label{friedm5} \end{equation}
with $\mpl^2=M^3r_c$.
This is the Friedmann equation in $f(\calrt)$ gravity with two new features:
a) An effective cosmological constant appears, despite the 
fine tuning of the bulk cosmological constant and the brane tension. 
b) The other striking feature is the negative sign of the contribution 
proportional to the energy density of the dark radiation.

For $\ex_2=-1$ and $kr_c \ll 1$ we have
\begin{eqnarray}
&&H^2+\frac{k_c}{R^2}+ 2\left[ 
(f'-1)\left( H^2 + \frac{k_c}{R^2}  \right)-\frac{1}{6}\left(\calrt f'-f\right)+H\dot{\calrt}f''
\right]=
\nonumber \\
&&\frac{4}{r_c^2}+
\frac{1}{6\mpl^2}\left(\rht-\frac{kr_c}{2}\rht_d \right),
\label{friedm6} \end{eqnarray}
with $\mpl^2=M^3r_c/2$.

It is obvious from the above that the brane cosmological expansion 
in the branch with $\ex_2=-1$ has novel properties arising
from:
a) an effective cosmological constant,
b) an effective negative energy density associated with the generalized
dark radiation, and
c) effective $f(\calrt)$ terms, that persist even in the limit $r_c\to 0$.
The second feature is not a consequence of the violation of
the weak energy condition, as the energy density is assumed positive both in
the bulk and on the brane. 
The first two features also appear in the DGP model
\cite{DGP}, characterized by $\Lx=V=0$ and $f=\calrt$. 

In the case of an AdS-Schwarzschild bulk we have $\rht_d \sim R^{-4}$.
At late times the contribution from the dark radiation is 
subleading to the contribution from the brane matter 
$\rht \sim R^{-3}$.
On the other hand, if there is a non-trivial matter configuration in
the  bulk so that $\rht_d \sim R^{-n}$ with $n<3$, 
the cosmological constant, the $f(\calrt)$ corrections and the effective negative
energy density are leading effects.

The conservation equation (\ref{ConservationEquation1}) (equivalently (\ref{conss}))  
takes a particularly simple form in the low-energy limit in which the term $\sim k^2$ dominates in the rhs.
We obtain
\begin{equation}
\left[ \dot{\rhot} +3 H (\rhot+\tilde{p}) \right]
+\epsilon_2
\left[ \dot{\rhot}_d +3 H (\rhot_d+\tilde{p}_d) \right]=0.
\label{consss} \end{equation}
This the effective four-dimensional description of the energy exchange between the brane and the bulk.

\section{Discussion} 
The generalisation of the DGP model with a $f(\calrt)$ term in the 4D action generates novel features 
in the brane cosmological evolution. An important question concerns the presence of regimes of accelerating expansion.
The inclusion of higher-derivative terms in the gravitational action may lead to such regimes. A typical example is 
the Starobinsky model in four dimensions \cite{starobinsky}. 
The evolution equations that we derived incorporate the Starobisnky model and its generalizations. 
The main behaviour is more transparent if we neglect the contributions from the matter sectors and the 
spatial curvature. Then, the effective Friedmann equation (\ref{friedm3}) at low energies becomes
\begin{equation}
H^2+ \frac{1-\ex_2+kr_c}{1+kr_c}\left[ 
(f'-1) H^2 -\frac{1}{6}\left(\calrt f'-f\right)+H\dot{\calrt}f''
\right]=
\frac{2(1-\ex_2)(1+kr_c)}{r_c^2},
\label{friedm3st} \end{equation}
with $\calrt=6(\dot{H}+2H^2)$.

The branch with $\ex_2=1$ leads to a generalization of the evolution in standard $f(\calrt)$ gravity, with the
higher derivative terms being multiplied by the factor $kr_c/(1+kr_c)$. It is easy to check that the evolution
equation admits accelerating solutions. For $f=\calrt+\calrt^2/(6E^2)$ we obtain behaviour similar to the 
Starobinsky model \cite{Felice, starobinsky}. There are 
approximate solutions of the form 
\begin{eqnarray}
&&H=H_i-\frac{\alpha E^2}{6}(t-t_i)+{\cal O}(E^4)
\label{st1} \\
&&\calrt=12H_i^2-\alpha E^2  +{\cal O}(E^4),
\label{st2} 
\end{eqnarray}
as long as the energy scale $E$ satisfies $H\gg E$, and 
$\alpha=({1+kr_c})/{k r_c}$.
For $kr_c \gg 1$, we have $\alpha=1$ and our setup reproduces exactly the 4D Starobinsky model. In the opposite limit
$kr_c \ll 1$, we have $\alpha \gg 1$ and the regime of accelerated expansion is much shorter than in the 
conventional scenario. 

The branch with $\ex_2=-1$ is the generalization of the self-accelerating branch of DGP cosmology
\cite{DGP,inducedrs,acceldgp,Apostolopoulos:2006si}.
The acceleration is induced by an effective cosmological constant 
$\sim (1+kr_c)/r^2_c$. If the higher derivative terms are neglected, 
the constant term in the rhs of equation (\ref{friedm3st})
leads to accelerated expansion with constant $H$ both for $kr_c \gg1$ and $kr_c \ll 1$. 
The second term in the lhs of equation (\ref{friedm3st}) generates terms proportional to higher powers of $H$. In general, accelerating
solutions are expected to exist even when the higher derivative terms are important. 
In the particular case of the Starobinsky model with $f=\calrt+\calrt^2/(6E^2)$, the higher derivative terms give a 
vanishing contribution to the self-accelerating solution. It must be emphasized that, within the standard DGP model, 
the self-accelerating branch has pathologies that render its physical signficance questionable 
 \cite {instability}. Similar problems may be present in the generalized version of the model that we are considering.
This issue will be the subject of future work. 

Our analysis, which was based on the use of covariant techniques, leads to a very intuitive understanding of the
effects of matter on the cosmological expansion at low energies. The brane matter contributes similarly to matter
in conventional 4D cosmologies. The effect of bulk matter can be incorporated in a term characterized as 
energy density of the generalized dark radiation. An appropriate pressure can be defined for this mirage component,
according to equation (\ref{darkpr}). In terms of these quantities the effect of the bulk on the cosmological evolution 
can be cast in a completely 4D form. A conservation equation can be derived (equation (\ref{consss})), describing the energy exchange
between the brane and the bulk, or, in other words, between the brane matter and the generalized dark radiation.
For the conventional branch with $\epsilon_2=1$, this equation indicates that the total energy of the brane and mirage matter sectors is
conserved, while the energy density is diluted by the expansion. For the self-accelerating branch with $\epsilon_2=-1$, 
equation (\ref{consss}) allows the energy of the two sectors to grow indefinitely. Another peculiar feature is the 
appearance of the contribution from the generalized dark radiation with a negative sign in the effective Friedmann
equation (\ref{friedm5}) or (\ref{friedm6}). This feature may lead to a cosmological evolution that crosses the 
phantom divide $w=-1$ \cite{Apostolopoulos:2006si}. The negative sign is not a consequence of the violation of any 
energy condition in the bulk-brane configuration, as both the bulk and brane matter components are assumed to have positive energy density.   

At early times and high energies, the behaviour becomes more complicated because of the presence of 
several unconventional contributions originating in the peculiarities of RS cosmology, the DGP model and $f(\calr)$ cosmology.
Despite the complexity of the setup, we have provided the equations that describe the general evolution.
They are the effective Friedmann equation (\ref{Friedmann}) or (\ref{friedm2}), and the effective conservation equation (\ref{ConservationEquation1}) or (\ref{conss}). 
They are applicable to all theories with an induced $f(\calrt)$ term and 
an arbitrary number of brane and bulk fields or matter components. They provide the basis 
for the analysis of the features of any specific model.

\vspace {0.5cm}
\noindent{\bf Acknowledgments}\\
\noindent
N.~T. is supported in part by the EU Marie Curie Network ``UniverseNet'' 
(MRTN--CT--2006--035863), the ITN network ``UNILHC'' (PITN-GA-2009-237920) and the program ```Kapodistrias'' of the University of Athens.

\end{document}